# Intrinsic amplitude-noise suppression in fiber lasers mode-locked with nonlinear amplifying loop mirrors

MARVIN EDELMANN,[1,2,3] YI HUA,[1,4,*] KEMAL ŞAFAK,[3] AND FRANZ X. KÄRTNER[1,4]

[1] *Center for Free-Electron Laser Science (CFEL), DESY, Notkestr. 85, 22607 Hamburg, Germany*
[2] *Department of Physics, Universität Oldenburg, Ammerländer Heerstr. 114-118, 26111 Oldenburg, Germany*
[3] *Cycle GmbH, Notkestr. 85, 22607 Hamburg, Germany*
[4] *Department of Physics, Universität Hamburg, Jungiusstr. 9, 20355 Hamburg, Germany*
*\*Corresponding author: franz.kaertner@desy.de*

**In this work, we investigate the steady-states of a fiber lasers mode-locked with a nonlinear amplifying loop-mirror that has an inherent amplitude noise-suppression mechanism. Due to the interaction of the sinusoidal transmission function with the fluctuating intracavity pulse amplitude we show that this mechanism may lead to a detectable difference in relative intensity noise at the reflected and transmitted output port under specific preconditions. We present systematic intensity noise measurements with a nonlinear fiber-based system that replicates a single roundtrip in the laser cavity. Experimental results and simulations clearly show a reduction of the intracavity amplitude fluctuations up to 4 dB for certain steady-states.**

Noise characteristics of mode-locked fiber lasers determine their suitability for a variety of cutting-edge applications in the field of synchronization and timing [1], frequency metrology [2], microwave generation [3], high speed optical sampling [4] and arbitrary waveform generation [5]. While those technologies are evolving at a tremendous speed, the requirements for reliable ultra-low-noise laser systems become increasingly severe and challenging. As a consequence, there has been an enormous effort in recent years to develop and optimize mode-locked sources for the generation of ultra-stable optical pulse trains. Especially, since applications for low-noise mode-locked laser has continuously grown outside specialized optical laboratories [6,7], fiber lasers mode-locked with asymmetric Sagnac-interferometers such as the nonlinear optical/ amplifying loop mirror (NOLM/ NALM) received a lot of attention in the last few years [8-11]. The reason behind this progression can be found in the advantages of these systems over other commonly used mode-locking devices and mechanisms. In comparison with widely used *real* saturable absorbers such as SESAMs [12] or two-dimensional nanomaterials (e.g. graphene and topological insulators) [13,14], the non-resonant artificial saturable absorber mechanism based on the optical Kerr-effect in NALM/NOLM mode-locked lasers is not restricted by carrier lifetimes in SESAM [15] and nanomaterials [16]. Furthermore, the possibility of an all-polarization-maintaining (PM) configuration is a major advantage over the well-known nonlinear polarization evolution mechanism greatly reducing sensitivity to environmental perturbations. In addition, lasers mode-locked with different implementations of the Sagnac-interferometer such as the commercialized Figure-9 fiber laser [17] have revealed record-low noise performance under the right conditions for reasons that have not yet been clarified. Considering that Sagnac-interferometers with a controlled degree of asymmetry have a long tradition in signal regeneration [18-21], reduction of amplitude-fluctuations [22,23] and generation of photon number squeezed light [24-26], it is reasonable to suppose that the superior low-noise performance of these mode-locked lasers is deeply connected with the inherent dynamics resulting from the interaction of the intracavity field with the self-amplitude modulation provided by the NALM.

In order to resolve the origin of this connection, we investigate the influence of the sinusoidal transmission function $T(\Delta\varphi_{nl})$ of a laser mode-locked with a phase biased NALM on the amplitude-noise (AM-noise) of the circulating intracavity pulse. To do so, we construct a novel NALM-based amplifier with an identically shaped transmission function $T(\Delta\varphi_{nl})$ and seed it with a commercially available ultra-low-noise short pulse laser. With this approach, we separate the influence of $T(\Delta\varphi_{nl})$ on the AM-noise from the optical feedback of the laser and enable the experimentally replication of one isolated roundtrip with identical physical mechanisms as they occur in the mode-locked steady state. Systematic measurements combined with numerical simulations clearly show, that the AM-noise of the circulating intracavity pulse per roundtrip is suppressed via the nonlinear transmission function, $T(\Delta\varphi_{nl})$, under the precondition, that the average nonlinear phase difference $\overline{\Delta\varphi_{nl}}$ corresponds to a working point on the transmission function, $T(\Delta\varphi_{nl})$, with a negative derivative.

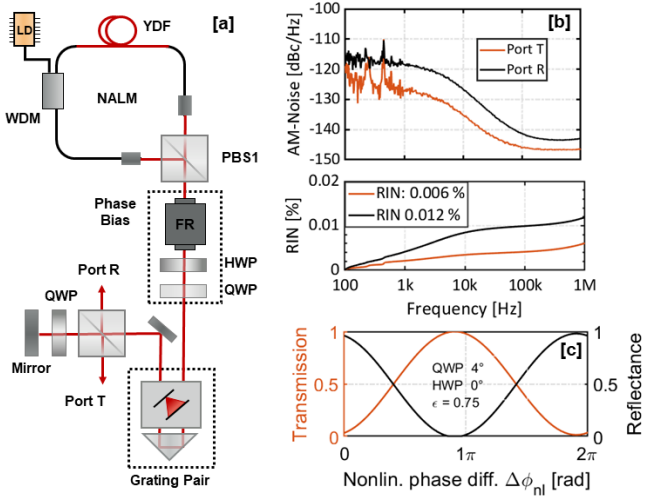

**Fig. 1.** [a]: Schematic of the all-PM NALM fiber laser. FR: Faraday-Rotator, WDM: Wavelength division multiplexer, LD: Laser diode, YDF: Ytterbium-doped fiber, HWP: Half-wave plate, QWP: Quarter-wave plate, PBS: Polarizing beam-splitter. [b]: Characteristic AM-Noise difference measured at Port T (orange) and Port R (black) with a corresponding RIN of 0.006% and 0.012 %, respectively. [c]: Corresponding transmission and reflectance functions for the specific mode-locked state with $\theta_{HWP} = 0°, \theta_{QWP} = 4°$ and an energy splitting ratio $\varepsilon = 0.75$ at the NALM entrance.

The NALM mode-locked laser used as a reference for the following measurements is shown in Fig.1 [a]. It is a modified, dispersion-managed Figure-9 laser with an all-PM NALM attached to a reflective arm. The asymmetric fiber loop is a sequence of 0.4-m single-mode fiber (SMF), 0.4-m highly Yb-doped gain fiber (CorActive Yb401-PM) and another 0.6-m segment of SMF. This inherent asymmetry in the fiber loop ensures the accumulation of a certain nonlinear phase difference $\Delta\varphi_{nl}$ even for a 50:50 energy splitting ratio at the entrance of the NALM. The fiber loop is attached to a reflective arm which contains the non-reciprocal phase bias (Faraday-rotator (FR), quarter waveplate (QWP) and half waveplate (HWP)) as well as an intracavity grating pair (1000 lines/mm) for tunable dispersion management. The repetition rate of the laser is ~80 MHz.

To characterize the AM-noise as precisely as possible in the following experiments, we detect the pulse train with a photodiode (ET-3000) and filter out the first harmonic of the received RF-signal. Further electronic amplification and usage of the built-in AM-noise measurement equipment of a signal-source analyzer delivers the frequency-resolved amplitude noise spectrum [27].

Fig.1 [b] shows the AM-noise of the reference NALM laser measured at the transmitted (Port T) and reflected port (Port R) for the same mode-locked state. The difference in integrated RIN (1kHz-1MHz) for Port T (0.006%) and Port R (0.012%) is roughly 3 dB. The corresponding state of $T(\Delta\varphi_{nl})$ as shown in Fig.1[c] ensures an energy splitting ratio of $\varepsilon = 0.75$ and a modulation depth of ~98% as a consequence of the particular phase bias. In fact, the AM-noise difference between both output ports is detectable over a wide range of adjustable cavity parameters (i.e. NALM pump power, net dispersion and phase bias), with the AM-noise of port T being either nearly equal or significantly lower than that of port R as it is also reported in Ref. [11].

The reason for this phenomenon can be found in the fundamental operation principle of asymmetric Sagnac-interferometers. The nonlinear phase shift for each counterpropagating pulse in the NALM depends on the gain g in the amplifier, the energy splitting ratio $\varepsilon$ of the counterpropagating pulses, the effective mode area $A_{eff}$ and the effective length $L$ of the nonlinear medium. It can be calculated using the relation

$$\delta\varphi_{nl,cw} = \frac{\pi}{\lambda_s A_{eff}} n_2 g L_{cw} P(1-\varepsilon) \quad (1)$$

for the clockwise propagating pulse and

$$\delta\varphi_{nl,ccw} = \frac{\pi L}{\lambda_s A_{eff}} n_2 g L_{ccw} P\varepsilon \quad (2)$$

for the counterclockwise propagating pulse in the nonlinear fiber loop under the assumption that both pulses experience similar gain in the Ytterbium-doped fiber (YDF) [28]. The transmission of the cavity depends on the intensity-dependent nonlinear phase difference $\Delta\varphi_{nl} = \delta\varphi_{nl,cw} - \delta\varphi_{nl,ccw}$, expressible through the transmission function $T(\Delta\varphi_{nl})$ which can be derived through an analysis of the cavity with the Jones-formalism [30]. Including a certain fluctuating intracavity peak power $P = \bar{P} + \delta P(t)$ resulting from the accumulated amplitude noise each roundtrip leads to the equation

$$\overline{\Delta\varphi_{nl}} + \delta\Delta\varphi_{nl}(t) = \frac{\pi}{\lambda_s A_{eff}} n_2 g(\bar{P} + \delta P[t])(L_{cw}[1-\varepsilon] - L_{ccw}\varepsilon) \quad (3)$$

which describes the transfer of peak power fluctuations into corresponding fluctuations of $\Delta\varphi_{nl}$ [29]. A further substitution of Eq. (3) into $T(\Delta\varphi_{nl}) = T(\bar{P} + \delta P[t])$ displays the cavity transmission as a function of the intracavity peak power which can be interpreted as a noise transfer function for the intracavity pulse amplitude fluctuations to the output ports. For the NALM laser, this noise transfer enables a suppression of the accumulated amplitude noise each roundtrip due to the inversely proportional fluctuation of the transmission to the peak power fluctuations for a negative derivative of $T(\bar{P} + \delta P[t])$. Simultaneously, the inverse transfer function $R(\bar{P} + \delta P[t])$ for the reflected port dictates an inverse nature of the measurable noise transfer, resulting in an amplification of the amplitude fluctuations for the reflected pulse.

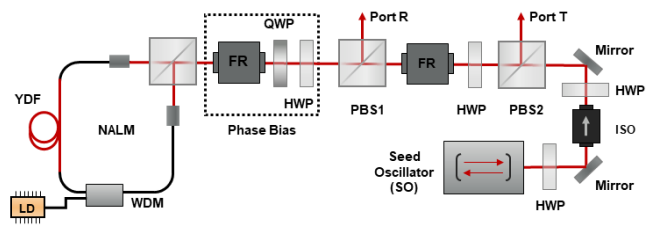

**Fig. 2.** Experimental setup of the NALM-based amplifier with the artificial transmission function. ISO: Isolator, SO: Seed oscillator

Experimentally, a systematic investigation of this inverse noise transfer to the output ports is not possible in the NALM-laser since the stability of mode-locked steady-state shows a high sensitivity against perturbations of the tunable cavity parameters. Hence, a NALM-based amplifier is constructed as shown in Fig. 2. Similar to the NALM-laser, this system consists out of an asymmetric all-PM NALM with 0.7-m SMF, 0.3-m highly doped YDF and another 1.5-m SMF in a sequence. The YDF is optically pumped with a laser diode emitting at 976 nm. The free-space arm attached to the fiber loop includes the non-reciprocal phase bias in form of a FR, HWP and a QWP. The linear polarized input signal is generated with a commercial seed oscillator which operates at 1030-nm center wavelength with 80-MHz repetition rate and a maximum average output power of 100 mW. The structure of this setup leads to a separation of the input pulse train into counterpropagating pulses in the NALM similar to the laser. PBS1 and PBS2 together with the FR and a HWP ensure that the separation of the reflected and transmitted component is based on the accumulated $\Delta\varphi_{nl}$ and the sinusoidal transmission function $T(\Delta\varphi_{nl})$ of the system. Therefore, it is possible to replicate an isolated roundtrip in the NALM laser with this nonlinear amplifier, since it physically modulates the low-noise input pulse train in an identical way to the modulation experienced by the intracavity pulse in a roundtrip of the NALM fiber laser. In the absence of optical feedback, such a NALM amplifier enables systematic parameter changes, which are necessary for an investigation of the noise transfer through $T(\Delta\varphi_{nl})$.

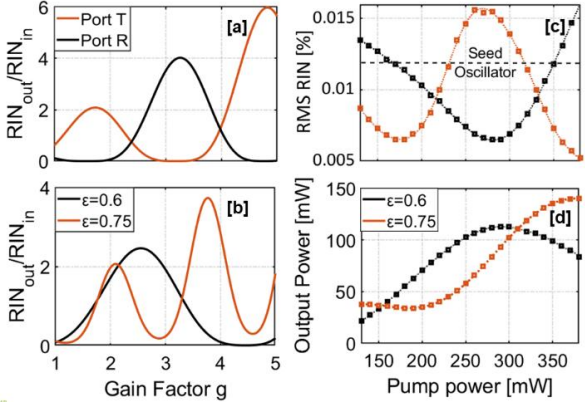

**Fig. 3.** [a]: Noise transfer of the NALM-based amplifier for $\bar{P} = 1 kW$, $\delta P = \pm 1\%$, $\varepsilon = 0.5$ and an asymmetric fiber loop ($L_{cw}$=1 m, $L_{ccw}$= 1.5 m) for Port T (orange) and Port R (black). [b]: Noise transfer at Port T for different degrees of asymmetry in the fiber configuration of the NALM. [c]: Measured noise transfer as a function of NALM YDF pump power measured at Port T with $\bar{P} = 2.2\ kW$ and an input RIN of 0.013 % from 1kHz to 1 MHz for different energy splitting ratios; $\varepsilon = 0.5$ (black) and $\varepsilon = 0.75$ (orange). [d]: Periodic modulated power transfer at Port T for similar parameters.

The important tuning parameters for the noise transfer are the gain in the NALM for a controlled change of $\Delta\varphi_{nl}$ as well as the rotation angles $\theta_{HWP}$ and $\theta_{QWP}$ in the phase shifter which are correlated with the offset and modulation depth of $T(\Delta\varphi_{nl})$. A simulation of the noise transfer for variations in the NALM gain can be seen in Fig. 3[a] for a fixed structural asymmetry ($L_{cw}/L_{ccw}$=2/3) and an energy splitting ratio of $\varepsilon = 0.5$. The simulation further shows the already mentioned inverse nature of the noise transfer for Port T and Port R in response to the shift of $\overline{\Delta\varphi_{nl}}$. The elongation of $\delta\Delta\varphi_{nl}$ increases proportional to the gain factor and results in more distinct changes of the transmission. Therefore, the magnitude of the noise transfer increases proportional to the gain factor. It should be noted, that the implied possibility for a complete amplitude noise suppression might be prevented through physical restrictions in the experiment such as shot noise and partial incoherent background radiation i.e. caused by amplified spontaneous emission generated in the YDF [31], which are not included in the simulation. Fig. 3[b] shows the computed noise transfer of the transmitted port influenced by different degrees of structural asymmetry in the NALM. As can be seen, the increasing asymmetry of the fiber loop leads to a higher periodicity of the noise transfer curve since the conversion of NALM gain to $\overline{\Delta\varphi_{nl}}$ becomes more efficient in agreement with Eq. (3). The experimental verification of this periodic noise transfer can be seen in Fig 3[c] for two distinct energy splitting ratios $\varepsilon$ at the NALM entrance. For this measurement, the rotation angle of the HWP is fixed at 0° and the QWP is rotated by 6° ($\varepsilon = 0.6$) and 112° ($\varepsilon = 0.75$). For both cases the RIN is then measured as a function of NALM YDF pump power. The reference RIN of the input pulse train generated by the seed oscillator is 0.013% (from 1kHz to 1MHz). Following the trend of the computed results, the measured noise transfer of the transmitted pulse train shows a periodic amplification and suppression of the input noise with respect to the reference value. For $\varepsilon = 0.75$, this noise suppression/amplification reaches a maximum value of ~2.5 dB and ~1.25 dB, respectively. A reduction of the energy splitting ratio from 0.75 to 0.5 changes the periodicity of the noise transfer function in agreement with the theoretical trend. Fig. 3[d] shows the output power measured at Port T as a function of the pump power. The periodic structure in the power transfer is a consequence of the periodically changing transmission for an increasing $\overline{\Delta\varphi_{nl}}$.

Another approach for tuning the noise transfer of the output ports includes the shift of $T(\Delta\varphi_{nl})$ for a fixed value of $\overline{\Delta\varphi_{nl}}$ and a fixed elongation of $\delta\Delta\varphi_{nl}$. This can be achieved with an adjustment of the phase bias between the cw and ccw propagating pulses in the NALM. In the experiment, this shift of $T(\Delta\varphi_{nl})$ corresponds to a fixed rotation angle $\theta_{QWP} = 0°$ and a variation of $\theta_{HWP}$.

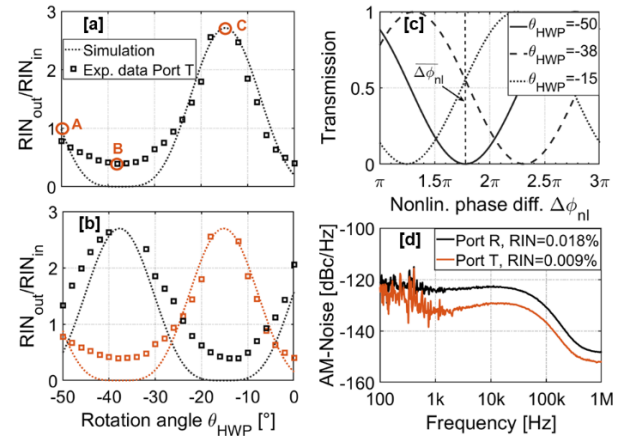

**Fig. 4.** [a]: Experimental results (marker) and simulation (line) of the noise transfer factor $\mathrm{RIN_{out}}/\mathrm{RIN_{in}}$ as a function of HWP rotation angle $\theta_{HWP}$. The input power of the seed is 30 mW with $\varepsilon = 0.5$, the NALM pump power is 450 mW. A, B, C: points with unchanged noise, max. reduction and max. amplification, respectively. [b]: Inverse noise transfer of Port T (orange) and Port R (black). [c]: Calculated $\overline{\Delta\varphi_{nl}}$ for dataset of [a] and [b] together with $T(\Delta\varphi_{nl})$ for states A, B and C in [a] with $\theta_{HWP}$=-50°, -38° and -15°, respectively [d]: Frequency-resolved amplitude-noise with 3dB noise difference of RIN integrated from 100 kHz to 1 MHz for Port T (orange) and Port R (black).

Fig. 4[a] shows the resulting periodic noise transfer curve, measured at port T with the maximum available NALM pump power of 450 mW. The integrated output noise RIN$_{out}$ (1kHz-1MHz) measured at Port T is shown normalized to the noise of the input pulse train (RIN$_{in}$= 0.013%) as a function of the rotation angle $\theta_{HWP}$. Three characteristic working points are marked with the maximum suppression/amplification of RIN$_{in}$ (A, C) as well as the point where it nearly remains unchanged (B). Fig. 4[b] shows the corresponding state of $T(\overline{\Delta\varphi_{nl}})$ for each of the marked points together with the calculated value of $\overline{\Delta\varphi_{nl}}$ =-8.7 rad normalized to $T(\overline{\Delta\varphi_{nl}})$ in the range of 0 to $2\pi$. In good agreement with simulations, it can be seen that the derivative of $T(\overline{\Delta\varphi_{nl}})$ determines the efficiency of noise transfer. The largest noise reduction (point B, $RIN_{out} = 0.4*RIN_{in}$) corresponds to the largest negative derivative of $T(\overline{\Delta\varphi_{nl}})$ and the highest noise amplification (point C, $RIN_{out} = 2.65*RIN_{in}$) corresponds to the highest positive derivative.

A comparison of the noise transfer output measured at port T and port R can be seen in Fig. 4[c] with a maximum noise difference of ~8.4 dB. Besides the inverse character of both curves as it was discussed already, a measurable noise difference at both output ports is inevitably connected with an amplification/suppression of the input noise RIN$_{in}$. For the case shown in Fig. 4[d] with a similar noise difference as in the laser (Fig. 1[a]) of ~3 dB, the noise transfer function shows that the noise of port T is simultaneously reduced by ~1.6 dB down to 0.009% integrated RIN (1kHz to 1 MHz) with respect to the input noise level.

In conclusion, we elucidated and quantified the amplitude noise suppression mechanism at work for the circulating intracavity pulses in NALM fiber oscillators per roundtrip in steady state. This has been achieved by studying the noise characteristics of a NALM mode-locked laser using a fiber amplifier mimicking the unfolded fiber laser cavity including the NALM. Based on experiments and simulation of the nonlinear amplifier, the origin of this effect is found to be the interaction of the NALM transmission function with the fluctuating intracavity pulse. Once the system operates in a regime where the derivative of the transmission function with respect to the pulse power negative, dynamic changes of the system transmission can reduce the amplitude noise of the transmitted pulse train. In the experiment, a noise reduction of up to 4 dB is realized by shifting $T(\overline{\Delta\varphi_{nl}})$ to the steepest negative slope. To confirm whether a particular NALM laser is mode-locked in such an amplitude-noise suppressed steady-state, one can detect the characteristic noise difference between the two output ports (i.e., port T lower than port R). This mechanism in combination with the high degree of tunability is a major advantage of lasers mode-locked with a NALM and explains the superior low-noise performance of NALM mode-locked lasers in general. The noise suppression mechanism detailed her can be used as a guide for the design of ultra-low noise fiber laser systems.